\documentclass[12pt]{article}

\usepackage[dvips]{geometry}

\begin{document}
\begin{large}
\textbf{Hans Albrecht Bethe} (July 2, 1906 -- March 6, 2005) 
\end{large}

\vspace{15pt}

German born American theoretical physicist whose reach and importance
in all fields of physics is towering. His work included: solid state
problems and the interaction of moving particles with matter (stopping
power); electrodynamical problems; nuclear physics; astrophysics;
meson theory; material physics; and hydrodynamics. 

In 1938 he published \emph{Energy Production in Stars} which explains the
CNO cycle that stars more massive than the sun use to fuse hydrogen to
helium in their cores. With Critchfield he also worked out the details
of the proton-proton process that is used as the main nuclear reaction
chain in the sun. Some of this work was done independently by
Weizsacker. In 1967 Bethe received the Nobel Prize for this work.
He was awarded the Max Planck Medal in 1955, and the Enrico Fermi Award in
1961. In 1944 he was elected to membership in the National Academy
of Sciences in Washington, DC and became a foreign member of the Royal
Society of London in 1957.

He is perhaps best known among nuclear physicists for publishing the
"Bethe Bible" a series of articles published in 1936 and 1937 in
\textsl{Reviews of Modern Physics} in collaboration with R.~F.~Bacher and
M.~S.~Livingston, written to explain the new theory of nuclear physics
to experimentalists. The review was actually a full textbook on the
field of nuclear physics which included many new calculations, some
performed with E.~J.~Konopinski and M.~E.~Rose. The series of articles
was reissued in 1986 under the title \emph{Basic Bethe}.

Bethe was born in Strassburg, Germany (now Strasbourg, France). His
father was a physiologist and his mother was a musician and writer of
children's plays. The Bethe family moved to Kiel in 1912 and again to
Frankfurt in 1915. He graduated from the Goethe Gymnasium in 1924 and
attended the University of Frankfurt from 1924-1926. At the age of 22
(1928) he received a Ph.D from the University of Munich were he worked
with A.~Sommerfeld, a wide-ranging theoretical physicist whose
students included W.~Pauli and W.~Heisenberg. Bethe accepted a
position in Stuttgart with P.~Ewald, and he would later marry Ewald's
daughter Rose.  They had two children, Henry and Monica, and three
grandchildren.

Bethe worked with Enrico Fermi as a Rockefeller fellow in Rome from
1931-1932. He returned to Germany and he published a review of
one-and-two electron problems in \textsl{Handbuch der Physik} and another which
deals with electrons in metals (solid state physics) with
Sommerfeld. In 1932 he obtained a position as a lecturer at the
University of Tuebingen until 1933. Because his mother was Jewish, he
became a refugee from the Nazis, going first to England where he
worked with R.~Peierls on the deuteron. He arrived in the US in 1935,
taking a position at Cornell University and remained there until his
death although he officially retired in 1975.

He played an important role in the war, leading the theory group at
Los Alamos. His initial war work came about after reading in the
Encyclopedia Britannica that the armor-piercing mechanism of grenades
was not understood; he formulated a theory that became the foundation
for research on the problem. After a stint at the Rad-Lab at MIT,
Bethe moved to Los Alamos where he played a crucial role in the design
of the atomic bomb. His war work made him an expert in shock waves and
hydrodynamics which he would later bring to bear on astrophysical
problems.

Bethe is best thought of explemplifying the scientist-statesman. He
was one of the founders of the \textsl{Bulletin of Atomic Scientists}, and he
donated a portion of his Nobel prize to help establish the Aspen
Center for Physics in Aspen, Colorado.  In 1958 he served as a member
of the US delegation to the first International Test Ban Conference in
Geneva. He wrote a very influential article in \textsl{Scientific American}
(with R.~Garwin) to which many attribute the adoption of the
Anti-Ballistic Missile Treaty (ABM Treaty).  During the 1980s, when
the treaty was threatened by the call for the development of nuclear
missile defenses (NMD), Bethe wrote numerous popular and technical
articles explaining why these would not work since any defense would
likely be overwhelmed by offensive decoys.  Bethe also worked for the
adoption of the Comprehensive Test Ban Treaty.

Bethe worked on the problems of re-entry of missiles and rockets into
the earth's atmosphere, an area which typically goes under the heading
of space-physics.

After his work on the energy production in stars, Bethe worked on
astrophysics problems off and on. In 1948 the article \emph{The Origin of
Chemical Elements} by R.~Alpher, H.~A.~Bethe, and G.~Gamow was
published in \textsl{Physical Review Letters}. Actually the work was done
by Alpher and Gamow, but Gamow so liked the sound of alpha,
beta, gamma, that he sent the paper in with the authors listed as
Alpher, Bethe (in abstentia), and Gamow. Bethe was the paper's
reviewer and his only change was to cross out the words "in
abstentia". In the 1960s Bethe worked on white dwarf stars with
R.~Marshak, and in the 1970s with the discovery of neutron stars, he
turned his attention to developing a theory of nuclear matter and the
equation of state in neutron stars. This work led him to work on
applying the theory of nuclear matter to finite nuclei which in turn
led him to collaborate with G.~E.~Brown. In 1979, while visiting
Brown in Copenhagen, Brown suggested they study the explosion
mechanism of "core-collapse" supernovae. That is the final fate of
massive stars which can fuse elements all the way up to iron in their
cores. These stars collapse and produce neutron stars or black holes,
but they also expel the heavy elements made during their lifetime into
the interstellar medium (which allows planets and life to
exist). Bethe and his collaborators worked initially on the prompt
mechanism of core-collapse whereby the incompressibility of nuclear
matter creates a shock wave which expels the outer parts of the star
and leaves behind either a neutron star or a black hole. But Bethe
also wrote an influential article with J.~Wilson interpreting one of
Wilson's numerical calculations and establishing the delayed
mechanism, where the shock stalls due to lack of energy, but is
re-invigorated by the deposition of energy from neutrinos released by
the proto-neutron star and infalling matter. The exact mechanism of
core-collapse is still a subject of active research. During the 1980s
Bethe, Brown, J.~Cooperstein, and students would spend January at
either Caltech or Santa Barbara (to escape the harsh weather of
Ithaca). That month was spent doing physics, cooking, and hiking in
the nearby mountains. It is often said that Bethe had two passions:
physics and hiking. With Brown, Bethe turned his attention to the
formation of stellar black holes in the 1990s.

In 1934, with Peierls, Bethe wrote an article on the possibility of
observing free neutrinos. Not foreseeing that large neutrino fluxes
would be available at nuclear reactors and particle accelerators, they
were pessimistic about the chance of directly observing neutrinos
(Reines and Cowan observed free neutrinos in 1956, Reines won the
Nobel Prize in 1995). However, Bethe particularly followed the work on
observing neutrinos from the sun that was pioneered by Ray Davis in
the late 1960s. Recognizing an important paper of Mikheyev and Smirnov
in 1986, Bethe wrote a \textsl{Physical Review Letter} showing how the work of
Mikheyev and Smirnov (and previous work of L.~Wolfenstein) could
reconcile the theoretical predictions for a larger neutrino flux than
was observed by Davis and later neutrino detectors. The most recent
observations at the Sudbury Neutrino Observatory in 2001 has shown this
interpretation to be correct.

Bethe is responsible for training nearly four generations of
physicists in solid state, atomic, nuclear, particle, and
astrophysics, an immense legacy.

\vfill\eject

\begin{large}
\leftline{\textbf{References:}}
\end{large}

\vspace{5pt}

\par\noindent\hangindent=10pt\hangafter=1
\emph{Perspectives in Modern Physics, Essays in Honor of Hans A.~Bethe},
R.~E.~Marshak and J.~W.~Blaker, eds., (New York: Wiley Interscience)
1966.

\vspace{5pt}

\par\noindent\hangindent=10pt\hangafter=1
\emph{Prophet of Energy: Hans Bethe}, Jeremy Bernstein, (New York:
Elesevier-Dutton) 1981.

\vspace{5pt}

\par\noindent\hangindent=10pt\hangafter=1
\emph{From a Life in Physics}, H.~A.~Bethe et al., (Singapore: World Scientific) 1989.

\vspace{5pt}

\par\noindent\hangindent=10pt\hangafter=1
\emph{Selected Works of Hans A.~Bethe With Commentary}, World Scientifc
Series in 20th Century Physics, Vol.~18, Hans A.~Bethe, (Singapore:
World Scientific) 1996.

\begin{flushleft}
Eddie Baron\\
Homer L.~Dodge Dept.~of Physics and Astronomy\\
University of Oklahoma\\
440 W.~Brooks, Rm 100\\
Norman, OK 73019-2061 USA
\end{flushleft}

\end{document}